**Concordia University**
**Department Of Computer Science and Software Engineering**

# LXG Compiler – Design and Implementation

by
Emil Vassev

December 16, 2003

# Table of contents

# 1. LXG Language Specifications

The LXG language is a simple language similar to Pascal by its semantic. Despite of the fact there is no practical use of the language, LXG is quite functional. It supports procedure and variable declarations, but no classes. The procedure's structure allows parameter passing by reference and value. LXG supports Integer, String and Boolean variable types. The arrays could be only one-dimensional of type Integer. In order to use a procedure, array or variable they must be declared first.

The control statements include IF-THEN, IF-THEN-ELSE, WHILE-DO and FOR-DO. The arithmetic operations are allowed only on integers and include *addition*, *subtraction*, *multiplication*, *division*, *division with reminder*, *raise to power* and *negation*. Boolean expressions may appear only in control and Boolean statements. Appendix A presents the LXG grammar in BNF (Backus Naur Form) format [1].

## 1.1. List of LXG token codes

### 1.1.1. Single-character symbol-tokens

| Token name | Lexeme |
|------------|--------|
| PLUS | + |
| MINUS | - |
| MULTI | * |
| OVER | / |
| POWER | ^ |
| COMMA | , |
| SEMI | ; |
| LESS | < |
| BIGGER | > |
| EQUAL | = |
| DIFF | # |
| LPAREN | ( |
| RPAREN | ) |
| LSQPAREN | [ |
| RSQPAREN | ] |

### 1.1.2. Multiple-character symbol-tokens

| Token name | Lexeme |
|------------|--------|
| ASSIGN | := |
| SWAP | :=: |
| LESSEQ | <= |
| BIGEQ | >= |
| ELLIPSIS | ,...., |

### 1.1.3. Reserved words

| Token name |
| --- |
| AND, DO, IF, OR, THEN, ARRAY, ELSE, INTEGER, PROCEDURE, VALUE, BEGIN, END, MOD, REM, WHILE, BOOLEAN, FOR, NOT, STRING |

### 1.1.4. Other tokens

| Token name | Description |
| --- | --- |
| ID | Identifier |
| NUMBER | Number |
| STRING | String |
| BOOLEAN | Boolean |

### 1.1.5. Bookkeeping

| Bookkeeping words | Description |
| --- | --- |
| bof | Begin of the file |
| eof | End of the file |
| error | An error discovered by the scanner |
| reserved word | Identifies the token as a reserved word |
| lexeme | The lexeme of the token if any |
| size | The size of the token (for strings and numbers) |

## 1.2. LXG Lexical conventions

### 1.2.1. Letter

- Letter = A|…|Z;

### 1.2.2. Digit

- Digit = 0|…|9;

### 1.2.3. Identifier

- Begins with a letter only;
- Consists of letters and digits only;

- There is a maximum allowable identifier length of 50 symbols;

### 1.2.4. Number

- Consists of digits only;
- There is a maximum allowable number length of 10 digits;

### 1.2.5. String

- Delimited by a start and end symbols, called delimiters;
- A delimiter could be ' or ", but not both;
- The start delimiter determines the end delimiter – if the start symbol is ' the end symbol is ' as well and vice versa;
- Any ASCII character may appear in a string, except the current delimiter (' or ").
- There is a maximum allowable string length of 256 symbols;

### 1.2.6. Boolean

- Boolean = TRUE|FALSE;

### 1.2.7. Comment

- Delimited by { and };
- Any symbol other than } may appear within a comment;
- Comments are ignored by the scanner;

### 1.2.8. Predefined LXG tokens

- Reserved words – see 1.1.3;
- Multiple-character symbols – see 1.1.2;
- Single-character symbols - see 1.1.1.

### 1.2.9. White space

- As white space could be considered any space (" "), any Tab, any LF and CR symbols;
- The scanner uses the white spaces as terminal symbols for token scanning to separate the reserved words form the identifiers in some cases – like variable declarations;
- The scanner ignores the white spaces. Like comments they are not a part of token stream;

### 1.2.10. EOF

- EOF is generated by the LXGScanner symbol that signs the end of the source file;
- EOF is the terminal symbol for the scanning process at all;
- It acts as terminal symbol for the current token, which the scanner works on;

### 1.2.11. Terminal symbol

- A terminal symbol for a token could be any symbol, which cannot appear within it. The scanner terminates a token scanning at the moment it encounters a symbol unacceptable for the current token;
- Common terminal symbols for all the tokens are white spaces and EOF;

### 1.2.12. Illegal symbol

- A symbol is defined as illegal if it cannot be included to the current token the scanner works on, cannot be recognized as common terminal symbol (white spaces or EOF) and cannot be recognized as start symbol for a new token or comment;
- Lower case characters are illegal outside of strings and comments;

### 1.2.13. Error

- Error is any terminated for scanning token, which is not recognizable as a LXG token;
- An illegal symbol is an error as well;

## 2. Design and Implementation

The Compiler for LXG Programming Language (LXG Compiler) has been implemented on Java ver. 1.4.2. For proper testing and execution, Java ver. 1.4.2. must be installed on the test machine.

The design and implementation of LXG Compiler has been based on OOP. The use of Java as a platform for implementation and the specification of the problem, mainly determined the use of OOP. The following class diagram (see Fig.1) illustrates the classes and relationships between them used in the LXG Compiler design and implementation. The class unit of each component of the system is divided into two sections – methods and class members. The class diagram shows no inheritance relations between the classes. The analysis of the problem has pointed to use of the Aggregation OOP model as more appropriate than the Inheritance OOP model. The classes (see Fig.1) represent functional units, which interoperate each other. The class hierarchy here is functional and depends on Aggregation model – the most superior class depends on all the classes and vice versa.

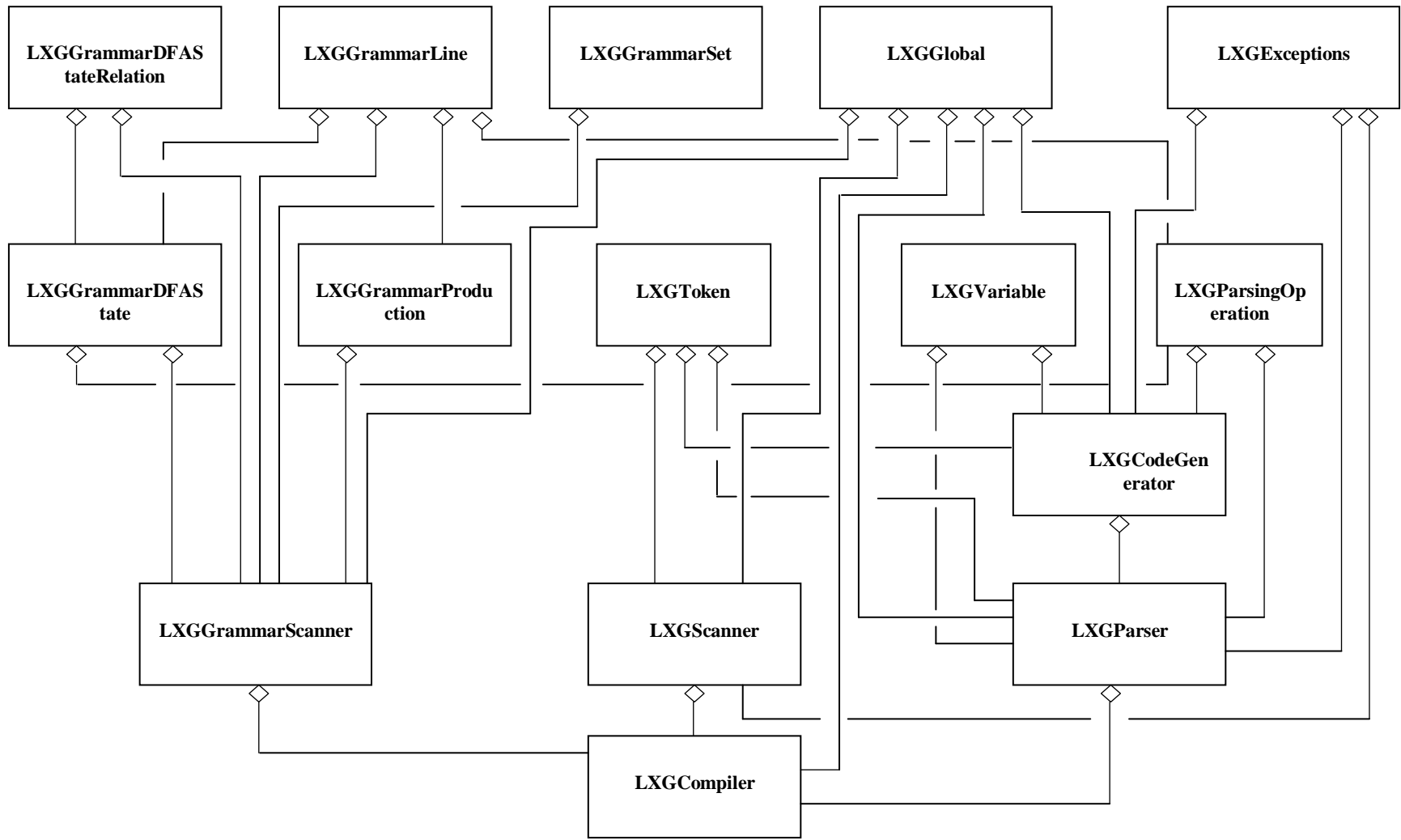

Fig.1. LXGCompiler class diagram

Fig.1 shows the Object model of LXG Compiler - designed and developed by me. This diagram describes the logical system entities, their classification and aggregation. Some of the classes represent the main steps in the compilation process - LXGScanner represents the Scanner, LXGParser represents the Pre-parser and Parser and LXGCodeGenerator represents the Code Generator of the Compiler (see Fig.2).

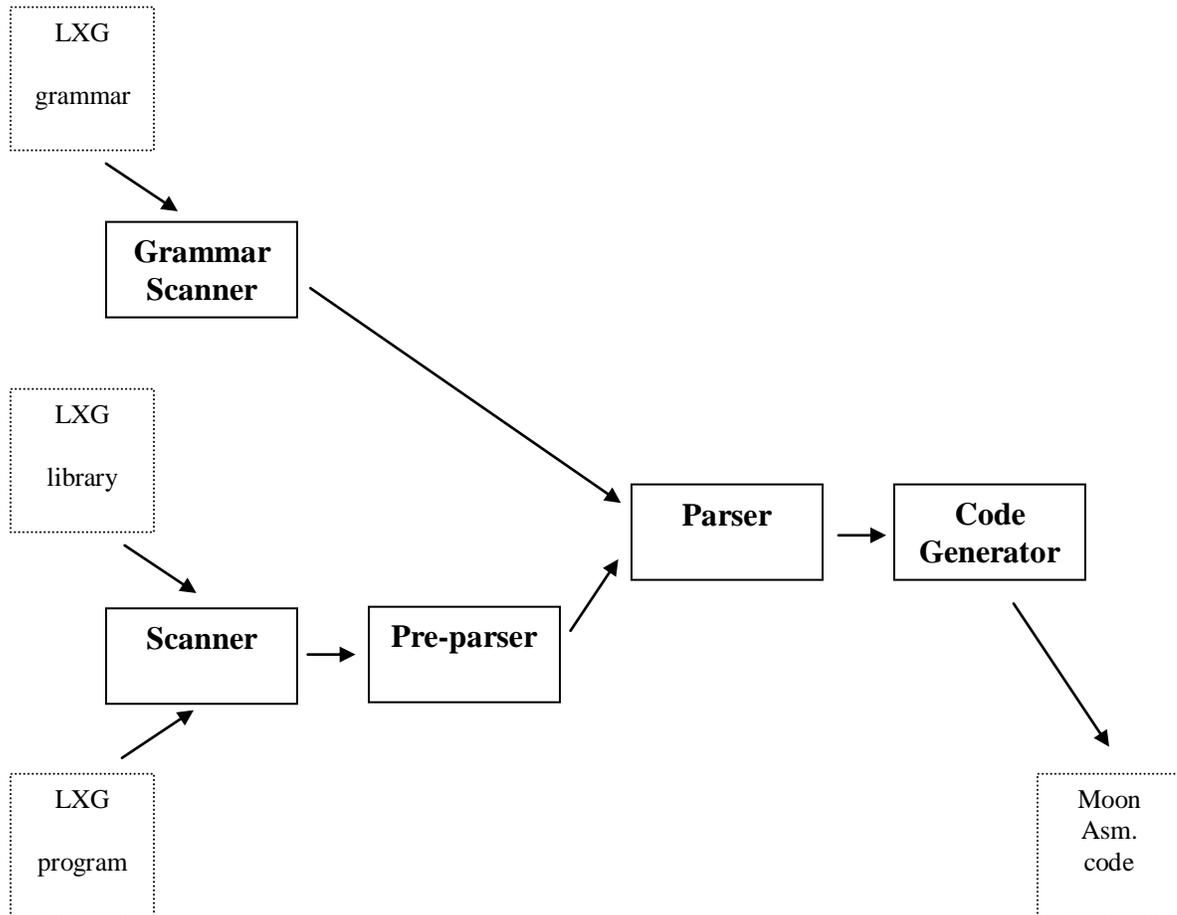

Fig.2. LXGCompiler unit block diagram

The LXG Compiler doesn't implement the LXG grammar rules hard-coded in the program. Instead it uses a special unit - Grammar Scanner for scanning the LXG grammar file and creates those rules dynamically. The class LXGGrammarScanner represents this functional unit (see Fig.1 and Fig.2).

The most superior class is LXGCompiler. It uses all the other classes in a direct or indirect manner.  With a few other classes – LXGGlobal, LXGExceptions and LXGToken, it forms the class framework. These classes are common for the project.

## 2.1. Class LXGCompiler

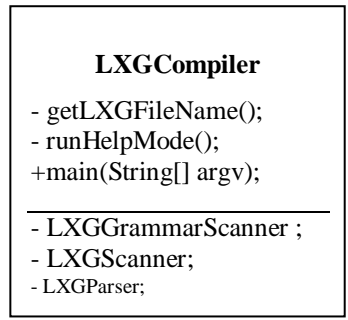

This is the main class of the LXG Compiler (see LXGCompiler.java). This class implements the ***main()*** function, which acts as an entry point to the program. This class has three methods - ***main()***, ***getLXGFileName()*** and ***runHelpMode()***, and three class members – instances of the classes ***LXGGrammarScanner***, ***LXGScanner*** and ***LXGParser***.

The ***main()*** method accepts as parameters the command line entries. If there is one, the program accepts it as the source file, which has to be compiled. If there is a second command line entry it should be "yes" or "no". The entry "yes" enforces the Compiler to dump to files all the middle results obtained in the compile process – grammar scanner out, scanner out, pre-parse out, symbol table and parse out. The code generator out is the final result of the process – an assembler file, appropriate to be executed on Moon machine. If the first parameter is a question mark "?" the LXG Compiler runs in a Help mode – shows a help information and stops.

Examples:
*java LXGCompiler test06.lxg*
*java LXGCompiler test06.lxg yes*
*java LXGCompiler ?*

If no command line entry is provided, the LXGCompiler calls the method ***getLXGFileName()***. This method prompts the user to enter the file name to be scanned. If the file name is defined LXGCompiler performs in a sequential order the following operations:

- It uses the ***LXGGrammarScanner*** class member to create an instance of the class and perform scanning on the LXG grammar file.
- It uses the ***LXGScanner*** class member to create an instance of LXGScanner and perform the scanning process on the source file.
- It uses the ***LXGParser*** class member to create an instance of LXGParser and perform the pre-parsing and parsing process on the source file. During the parsing process, the LXGParser uses the LXGCodeGenerator to generate the final assembler code.

The method ***runHelpMode()*** runs the help mode. It shows the help information.

## 2.2. Class LXGGrammarScanner

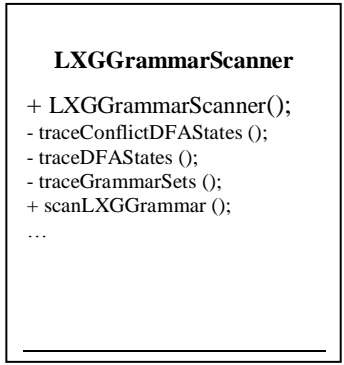

**LXGGrammarScanner**

+ LXGGrammarScanner();
- traceConflictDFAStates ();
- traceDFAStates ();
- traceGrammarSets ();
+ scanLXGGrammar ();
…

The class of LXGGrammarScanner (see LXGGrammarScanner.java) implements all the functionalities necessary to perform the scanning process on the LXG grammar file, and create the set of grammar rules used by the other compiler's units. The class has few principle methods – ***scanLXGGrammar(), traceGrammarSets(), traceDFAStates(), traceConflictDFAStates()***, a bunch of helpful additional methods and some private class members, which are specific for the LXGGrammarScanner only.

The method ***scanLXGGrammar()*** triggers the scanning process. It opens the LXG grammar file, and performs in a sequential order the following operations:

- Calls repeatedly the method ***scanNextItem()*** to read the grammar file and fill up a grammar dynamic structure;
- Calls ***traceGrammarSets()*** to compute grammar's First and Follow sets [2];
- Calls ***traceDFAStates()*** to compute all the DFA [2, 3, 4, 5] states and keep them in the LXGGlobal.grammarDFAStates vector;

The following algorithm has been implemented for constructing the DFA states:

- Create a new nonterminal S' and a new production S' $\rightarrow$ S where S is the start symbol.
- Put the item S' $\rightarrow$ • S into a start state called state 0.
- Closure: If A $\rightarrow$ α•Bβ. is in state s, then add B $\rightarrow$ •### to state s for every production B $\rightarrow$ ### in the grammar.
- Creating a new state from an old state: Look for an item of form A $\rightarrow$ α•xβ. where x is a single terminal or nonterminal and build a new state from A $\rightarrow$ αx•β. Include in the new state all items with •x in the old state. A new state is created for each different x.
- Repeat steps 2 and 3 until no new states are created. A state is new if it is not identical to an old state.

The method ***traceDFAStates()*** calls at its end the method ***traceConflictDFAStates()*** for determining the shift-reduce conflicts.

When the Grammar Scanner is done we have:

- The LXG grammar loaded in a special format (see LXGGrammarLine class) and kept in the LXGGlobal.grammarLines vector;
- All the DFA states are created in a special format (see LXGGrammarDFAState), and kept in LXGGlobal.grammarDFAStates vector;
- All the First and Follow sets are created in a special format (see LXGGrammarSet class) and kept in LXGGlobal.firstSets and LXGGlobal.followSets vectors.

If the LXG Compiler has been run into a "dump to files" mode (see 2.1) the Grammar Scanner runs the method ***writeDFAStatesToFile()*** to save all the DFA states with their Reduce, Shift and Goto transitions.

### 2.3. Class LXGScanner

```
┌─────────────────────────────────────┐
│           LXGScanner                 │
├─────────────────────────────────────┤
│ + LXGScanner();                      │
│ - boolean isLetter();                │
│ - boolean isDigit();                 │
│ - boolean isSingleToken();           │
│ - boolean isReservedWord();          │
│ - boolean isBoolean();               │
│ …                                    │
│ - writeScannedToken();               │
│ - boolean scanNextToken();           │
│ + scanLXGFile();                     │
└─────────────────────────────────────┘
```

The class of LXGScanner (see LXGScanner.java) implements all the functionalities necessary to perform the scanning process. The class has few principle methods – ***scanLXGFile(), scanNextToken(), writeScannedToken(),*** a bunch of helpful additional methods and some private class members, which are specific for the LXGScanner only.

The method ***scanLXGFile()*** triggers the scanning process. It scans first the LXG library file and after that scans the LXG source file - opens the source file, creates the output file and calls repeatedly the method ***scanNextToken()***. When ***scanNextToken()*** returns false, scanLXGFile() closes all the open files and returns to its caller – ***LXGCompiler.main()***.

The method ***scanNextToken()*** scans the source file for the next possible token. A sophisticated algorithm is used to recognize a token and to determine the token type. The algorithm uses the declarations from LXGGlobal. If a token is found, ***scanNextToken()*** calls the method ***writeScannedToken()*** to write the token into the output file - the file for the stream of tokens. If an error is discovered ***scanNextToken()*** determines its kind and writes it down by calling again ***writeScannedToken().***

The method **writeScannedToken()** writes the current token into the output file. This includes also writing of the token's lexeme if any, and size (for NUMBER and STRING).

When the LXG Scanner is done we have all the scanned tokens loaded in a special format (see LXGToken class) and kept in two dynamic vectors:
- LXGGlobal.tokenLibraryTable – keeps the library file tokens;
- LXGGlobal.tokenTable – keeps the source file tokens.

The LXG Scanner provides an error-trapping mechanism, which catches incorrect symbols, incorrect tokens, incorrect string size (max 256 symbols), incorrect identifier size (max 50 symbols) and incorrect number size (max 10 digits) errors.

### 2.4. Class LXGParser

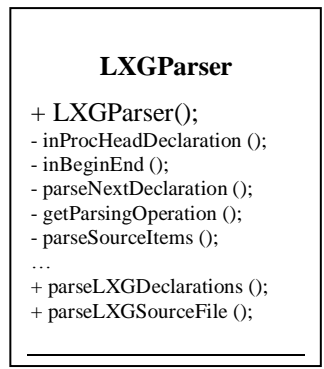

The class of LXGParser (see LXGParser.java) implements two main units of the LXG Compiler – Pre-parser and Parser (see Fig. 2). The Pre-parser parses all the variable and procedure declarations and creates the symbol table. The Parser uses the dynamic structures created by the Grammar Scanner, the token tables created by the LXG Scanner and the symbol table to perform the parsing process.

The LXGParser class has few principle methods – **parseLXGSourceFile(), parseLXGDeclarations(), parseSourceItems(), getParsingOperation(), parseNextDeclaration(),** a bunch of helpful additional methods and some private class members, which are specific for the LXGParser only.

#### 2.4.1. Pre-parser

The Pre-parser called also Declaration parser accepts as input the token stream produced by the scanner. The output is a modified token sequence, with variable and parameter declarations (including semicolons) removed and identifier tokens replaced as explained below. In addition, the declaration parser produces from all the declarations, including the procedures, a dynamic structure called Symbol table (or Variable Declaration table) (see LXGVariable class).

The following are the rules used by the Pre-parser to construct the Symbol table and to create a new token output:

- An identifier token appearing outside of a variable declaration is replaced by one of iIdentifier, bIdentifier, sIdentifier, aIdentifier, uIdentifier as it is determined by the variable declaration table.
- The declaration parser identifies and retains the sequence **PROCEDURE uIdentifier (ident-list)**, until the types of the parameters in ident-list have been determined by subsequent declarations.
- The same variable name may be declared in the main program and in one or more procedures. Each such declaration defines a different variable, possibly with a different type. The pre-parser allows for the possibility that some variable declarations will be (temporarily) superseded by declarations within a procedure.

LXGCompiler.main() triggers the parsing declarations process by calling the public method **parseLXGDeclarations()**. Inside, this method calls two times repeatedly **parseNextDeclaration()** method . First the method run to parse the LXG library file and second to parse the LXG source file.

The **parseNextDeclaration()** method does sequentially the following actions:
Analyses where the current token is – in a procedure or in the main program;

- Calls **inDeclaration()** method if the token is within a variable declaration;
- Calls **inBeginEnd()** method if the token is within BEGIN-END block;
- Calls **inProcHeadDeclaration()** method if the token is within a procedure head declaration;
- Calls **setVarTokenType()** method to set the token type - PROCEDURE or a variable type (STRING, BOOLEAN, INTEGER, ARRAY);
- Calls **writeVarInSymbolTable()** if the token is recognized as a variable or procedure to save it in the symbol table.

The **inDeclaration()** method analyses the variable declarations and add them to the symbol table. If the declaration is an ARRAY the method **inArrayDeclaration()** is called. After saving the variables into the symbol table by calling the **writeToSymbolTable()** method, the **inDeclaration()** method deletes the declaration section from the token stream by calling **deleteFromTokenTable()** method.

The following errors could be trapped by the **inDeclaration()** or **inArrayDeclaration()** methods:

- Incorrect declaration – unexpected token;
- Incorrect array declaration - the array size must be greater than zero.

The **inProcHeadDeclaration()** method analyses procedure head declarations and add the parameters to the symbol table. The type of those parameters is set up later on when their declaration is found.

All the three methods described above – inDeclaration(), inArrayDeclaration() and inProcHeadDeclaration(), call the **correctVarDeclaration()** method  to determine the correctness of the variable or procedure declaration.   A correct declaration is not:

- A duplicate declaration within the same location – procedure or the main program;
- A VALUE declaration of an array parameter;
- A VALUE declaration of a parameter, which is not declared;
- A VALUE declaration within the main program.

### 2.4.2. Parser

The Parser implemented by LXG Compiler is a SLR(1) parser. It is a Bottom-up parser. Bottom-up parsing methods have an advantage over top-down parsing in that they are less fussy in the grammars they can use. The LXG Parser uses the DFA Grammar states and Follow sets constructed by LXGGrammarScanner (see 2.2.). The Parse table is integrated in the DFA states.

The following describes the SLR(1) specifications implemented by LXG Parser:
- Resolve shift-reduce conflicts in favor of shifting;
- Reduce-reduce conflicts do not appear.
- Reduce rule: only reduce when next input symbol is an element of Follow set of the resulting nonterminal;
- Shift moves are associated with transitions on terminal symbols;
- Goto moves are associated with transitions on nonterminal symbols;
- Reduce moves are associated with complete items.

The LXG Parser accepts as an input a stream of tokens from the Declaration parser (Pre-parser), and as an output, the parser produces (in a reverse order) a sequence of LXG rules that give a rightmost derivation of the input string.

The Parser works together with the Code Generator (see LXGCodeGenerator class). Each reduce or shift parse operation calls the Code Generator in order to generate a code corresponding to that operation.

LXGCompiler.main() triggers the parsing process by calling the public method ***parseLXGSourceFile()***. This method calls ***parseSourceItems()*** method. Also, it creates the Code Generator object (instantiated from LXGCodeGenerator), and calls its ***LXGCodeGenerator.start()*** and ***LXGCodeGenerator.stop()*** methods.

The ***parseSourceItems()*** method performs the actual parsing process. This process involves the following steps in a sequential order:
- Creates an empty parse stack and push in the first DFA state:

Example:

  program $\rightarrow$ **.***bof* prgm-body ***eof***

- Reads the next token produced by the Declaration parser, by calling ***getNextToken()*** method;
- Calls repeatedly (in a loop) the getParsingOperation() method, to get the next parsing operation;
- If the parse operation is Shift:

- o It pushes into the parser stack the operation's next state (see LXGParsingOperation class);
- o It reads the next token from the token stream by using ***getNextToken()*** method;
- o It calls ***LXGCodeGenerator.generateCodeTags()*** method to generate the necessary code;
- If the parse operation is Reduce:
  - o It calls ***LXGCodeGenerator.generateCode()*** method to generate the necessary code;
  - o It pops from the parser stack the reduced items;
- If the parse operation is Reduce and the reducing is to the first grammar production this is the end of parsing process – breaks the loop.
  Example:

  program $\rightarrow$ ***bof*** prgm-body ***eof.***

The key for the process described above is the determination of the next parsing operation performed by ***getParsingOperation()*** method. The method accepts as parameters the DFA state and token. The DFA states contains all the shift, reduce and goto transition, concerning this state (this is the parsing table for the DFA state). The ***getParsingOperation()*** method determines the next DFA state by using those transitions and construct the next parsing operation. If there is a shift-reduce conflict it is resolved in favor of shifting.

The Parser traps the following errors:
- Any syntax error – any LXG grammar illegal syntax;
- Any conflict between the argument types in a procedure call and the parameter types of the procedure.

## 2.5. Class LXGCodeGenerator

```
LXGCodeGenerator

+ LXGCodeGenerator();
- stop ();
+ start ();
- writeArrOp ();
- writeErrorTraps ();
- newDenominator ();
…
+ generateCode ();
+ generateCodeTags ();
```

The Code Generator unit (see LXGCodeGenerator.java) is the last compiler unit, which finalizes the work done by the compiler – generate the assembler code. Since no code optimization is required, the generator generates the code for the Moon machine without any use of intermediate code or intermediate representation. This unit works together with the Parser – in any parser pass there is a code generation.

The generated code is compatible with the Moon processor. The Moon assembly language is simple but efficient. The processor has a few instructions that access memory and many instructions that perform operations on the contents of registers. Since register access is much faster than memory access, the Code Generator uses those registers efficiently. The Moon Processor has sixteen registers, R0 - R15 and no stack. The Code Generator maintains the registers as follow:

- R0 is used as an IP (instruction pointer);
- R15 is used to maintain a program stack;
- R14 is used as a link register;

Example:

> ***jl    R14,PGEN***

- In any procedure call the Code Generator save all the registers at the entry point of the procedure and restore the registers' values at the end of the procedure.

The LXGCodeGenerator class has few principle methods – ***generateCode(), generateCodeTags(), newDenominator(), writeErrorTraps(), writeDecalarations(), start(), stop(),*** a bunch of helpful additional methods and some private class members, which are specific for the LXGCodeGenerator only.

The first method needed to be executed is the ***start()*** method. This is the initial entry point to the Code Generator. It creates an empty out file and writes the header information into the file.

The actual code generation process is performed mainly by ***generateCode()*** and ***generateCodeTags()*** methods. Those two methods are run repeatedly by the parser's method ***parseSourceItems()*** (see 2.4.2.).

The ***generateCode()*** method is run at any Reduce Parser action and generates the appropriate code. As parameters the method receives the current parsing operation (see 2.4.2.) and token. A complex internal mechanism determines what code should be generated. Any intermediate generated code is pushed into the ***codeStack*** stack variable. The correct use of registers is implemented by tracking, which registers are currently in use, and by using only the free ones. This technique is implemented by ***freeRegister()*** method. To reduce complexity and redundancy the ***generateCode()*** method uses a bunch of help methods like ***newDenominator(), doLoop(),    addToDest(), opDelete(), opPopArgCheck(), opPopRegCheck(), writeArrOp(), asVariable()*** and ***asRegister()***.

The ***generateCodeTags()*** method is run at any Shift Parser action and generates the appropriate code. As a parameter the method receives the current token (see 2.4.2.). This method helps in code generation by generating some labels mainly for the control statements FOR-DO, WHILE-DO and IF-THEN. Both methods ***generateCodeTags()*** and ***generateCode()*** use a mechanism for label and variable name generation. This mechanism is provided by the methods ***createGenVariable()*** and ***getGeneratedName()***. Also, the ***generateCodeTags()*** method determines when to generate the code corresponding to the entry point for the program. The code generation of the entry poin is performed by the *writeEntryPoint()* method. The entry point should be after all the procedure declarations. The first operation after the entry point is the initialization of the stack pointer (R15).

Example:

*%----------------- program's entry point ------------------%*
            *entry*

            *jl    R14,Sy_Init_SP*

The code generation for FOR-DO statements requires computation of the denominator at any change for-list change. This is implemented by the method *newDenominator()*. The method helps the *generateCode()* method by generating the code corresponding to the denominator computation.

Example:

 *LB_00005*
*%---- denominator ---%*
            *lw    R1,B(R0)*
            *lw    R2,A(R0)*
            *sub   R3,R1,R2*
            *sw    I_00003(R0),R3*

*%----------------- zero denominator check ------------------%*
            *bz    R3,LB_FOR_DO_ZERO_DEN*

At the end of the code generation LXGCodeGenerator calls the *writeErrorTraps()* method. This method writes all the run-time error traps as follow:

- Recursive function call error – the compiler does not support recursive function calls;
- FOR-DO zero denominator error – the denominator must be different than zero;
- Array zero-index writing error – cannot write to the zero index element of an array;
- Array low boundary error – the index of an array cannot be less than 0;
- Array up boundary error - the index of an array cannot be bigger than the array's top boundary;
- Division by zero error;
- Negative exponent error.

Also, the *writeErrorTraps()* method at its end calls the *writeDecalarations()* method. This method is used to generate code for all the variables - generated or declared.

The final method executed by LXGCodeGenerator is the *stop()* method. This method closes the out file (the assembler file).

The LXG language is not a complex one, but maintains some complex structures like arrays, procedures, FOR-DO loops, WHILE-DO loops, IF-THEN and IF-THEN-ELSE control statements. The translation of those structures into Moon assembly language requires some complexity reduction. The following is a brief description of the code generation concerning those structures with applied examples.

### 2.5.1. Code Generation for LXG arrays

The array representation in Moon assembly language is a sequence of words, the first of which gives the size of the array. The code generated for an array declaration by the LXG Code Generator is:

LXG: *ARRAY A01[100]*

Assembler: *A01            dw    100*
*res    400*

Where, A01[0] is the array size.

### 2.5.2. Code Generation for IF-THEN and IF-THEN-ELSE statements

The control statements IF-THEN and IF-THEN-ELSE requires some label generation and internal jump statements to those labels. The code generated for a simple IF-THEN-ELSE statement is:

LXG:        *IF A>B  THEN WRITES("A>B")*
*ELSE WRITES("NOT A>B");*

Assembler:

```
            lw    R2,A(R0)
            lw    R3,B(R0)
            cgt   R1,R2,R3
            j     LB_00002
LB_00001
            jl    R14,Sy_Push
            dw    S_00001
            jl    R14,WRITES

            j     LB_00004
LB_00003
            jl    R14,Sy_Push
            dw    S_00002
            jl    R14,WRITES
```

```
              j    LB_00004
  LB_00002
%------------------ IF BOOL_EXP THEN ... ------------------%
              bnz   R1,LB_00001
%------------------ ELSE ... -----------------------------%
              j    LB_00003
  LB_00004
```

### 2.5.3. Code Generation for WHILE-DO statement

The control statement WHILE–DO requires some label generation and internal jump statements to those labels. The code generated for a simple WHILE-DO statement is:

LXG:  ***WHILE A>B DO WRITES("A>B")***

Assembler:
```
%---------------------- WHILE-DO: WHILE --------------------%
  LB_00001
              lw    R2,A(R0)
              lw    R3,B(R0)
              cgt   R1,R2,R3
              j    LB_00002

%----------------------- WHILE-DO: DO ----------------------%
  LB_00003
              jl    R14,Sy_Push
              dw    S_00001
              jl    R14,WRITES

              j    LB_00001
  LB_00002
%------------------ WHILE BOOL_EXP ... ------------------%
              bnz   R1,LB_00003
```

### 2.5.4. Code Generation for FOR-DO statement

The control statement FOR–DO requires some label generation and internal jump statements to those labels. This is the most complex statement and different ***for-list*** (see the LXG grammar FOR-DO statement) representations lead to different code generation. The code generated for a simple FOR-DO statement is:

LXG:      ***FOR A:= 1,...,8 DO WRITES("A")***

Assembler:

```
%---------------- BEGIN: FOR ... DO ...      ----------------%
            j    LB_00001
%------------------------ FOR-DO: DO ----------------------%
  LB_00002
            sw    REG_00000_LB_00002(R0),R14
            jl    R14,Sy_Push
            dw    S_00001
            jl    R14,WRITES

            lw    R14,REG_00000_LB_00002(R0)
            jr    R14

%---------------- FOR-DO: calculation ...    ----------------%
  LB_00001
            lw    R1,INT_VAL_ONE(R0)
            sw    I_00003(R0),R1
            lw    R1,I_00001(R0)
            sw    A(R0),R1
            lw    R1,I_00003(R0)
            cgti  R1,R1,0
            bz    R1,LB_00004

  LB_00003
            lw    R1,I_00002(R0)
            lw    R2,A(R0)
            cle   R2,R2,R1
            bz    R2,LB_00005
            jl    R14,LB_00002
            lw    R2,I_00003(R0)
            lw    R3,A(R0)
            add   R2,R3,R2
            sw    A(R0),R2
            j     LB_00003
  LB_00004
            lw    R1,I_00002(R0)
            lw    R2,A(R0)
            cge   R2,R2,R1
            bz    R2,LB_00005
            jl    R14,LB_00002
            lw    R2,I_00003(R0)
            lw    R3,A(R0)
            add   R2,R3,R2
            sw    A(R0),R2
            j     LB_00004
```

**LB_00005**
**%----------------- END: FOR ... DO ...        ---------------%**

### 2.6. Class LXGToken

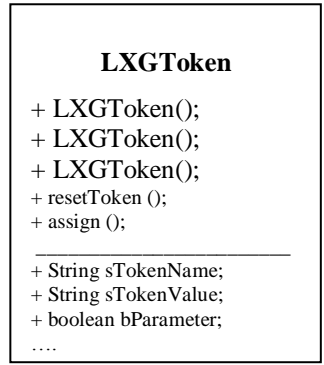

The class LXGToken (see LXGToken.java) represents the LXG token structure used by the LXG Compiler. All the methods and class members are public. It makes the use of the class easier. The class is for internal use only, and the public class members do not violate the class privacy and capsulation.

There are three overloaded constructors implemented within the LXGToken class. They differ by their parameters.

The method ***resetToken()*** resets all the class members to their initial state.

The method ***assign()*** assigns a LXGToken to the current one.

### 2.7. Class LXGVariable

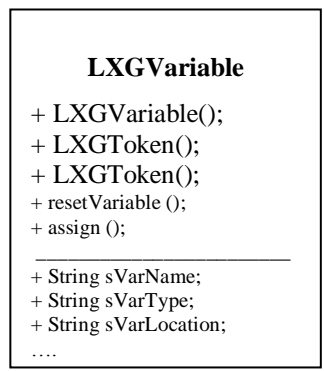

The class LXGVariable (see LXGVariable.java) represents the LXG variable structure used by the LXG Compiler. All the methods and class members are public. It makes the use of the class easier. The class is for internal use only, and the public class members do not violate the class privacy and capsulation.

There are three overloaded constructors implemented within the LXGVariable class. They differ by their parameters.

The method ***resetVariable()*** resets all the class members to their initial state.

The method ***assign()*** assigns a LXGVariable to the current one.

## 2.8. Class LXGGrammarDFAState

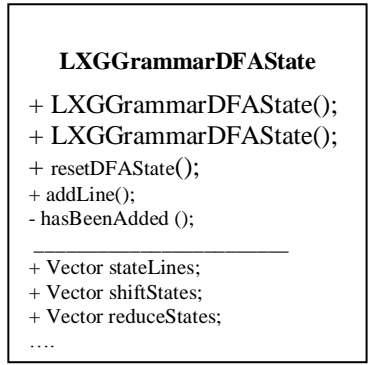

The class LXGGrammarDFAState (see LXGGrammarDFAState.java) represents the LXG grammar DFA structure used by the LXG Compiler. All the methods and class members are public except the method ***hasBeenAdded()***. This method checks if the current grammar line is been already added to the DFA state. The public class members make the use of the class easier. The class is for internal use only, and the public class members do not violate the class privacy and capsulation.

There are two overloaded constructors implemented within the LXGGrammarDFAState class. They differ by their parameters.

The method ***resetDFAState()*** resets all the class members to their initial state.

The method ***addLine()*** adds a new grammar line (see LXGGrammarLine class) to the current set of lines kept by the DFA state into the ***stateLiness*** vector.

## 2.9. Class LXGGrammarProduction

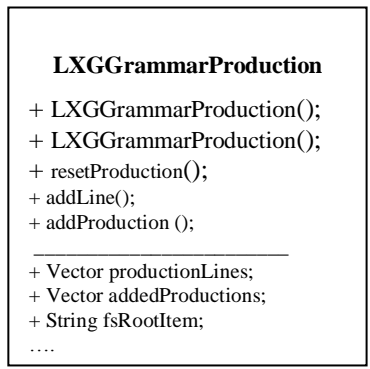

The class LXGGrammarProduction (see LXGGrammarProduction.java) represents the LXG grammar production structure used by the LXG Compiler. All the class members are public. The public class members make the use of the class easier. The class is for internal use only, and the public class members do not violate the class privacy and capsulation.

There are two overloaded constructors implemented within the LXGGrammarProduction class. They differ by their parameters.

The method **resetProduction()** resets all the class members to their initial state.

The method **addLine()** adds a new grammar line (see LXGGrammarLine class) to the current set of lines kept by the Grammar production into the **productionLines** vector.

The method **addProduction()** adds a Grammar production to the current set of added productions kept by the Grammar production into the **addedProductions** vector.

### 2.10. Class LXGParsingOperation

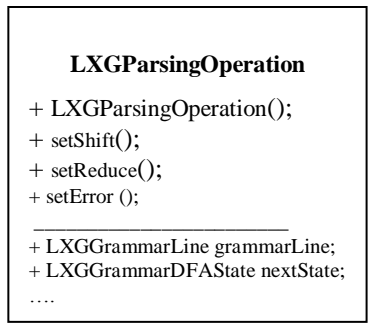

The class LXGParsingOperation (see LXGParsingOperation.java) represents the LXG parsing operation structure used by the Parser. All the class members are public. The public class members make the use of the class easier. The class is for internal use only, and the public class members do not violate the class privacy and capsulation.

There is one constructor implemented within the LXGParsingOperation class.

The method **setShift()** set the operation as a Shift parsing operation.

The method **setReduce()** set the operation as a Reduce parsing operation.

The method **setError()** set the class member **fbError** to true which used by the Parser to catch the syntax errors.

### 2.11. Class LXGGrammarDFAStateRelation

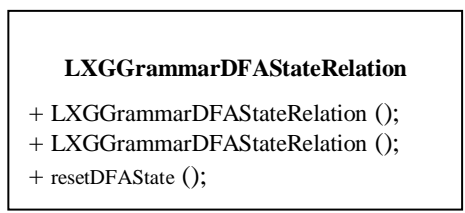

The class LXGGrammarDFAStateRelation (see LXGGrammarDFAStateRelation.java) represents the LXG Grammar DFA state relation structure used by the Grammar scanner unit. There are no other class members but three public methods.

There are two constructors implemented within the LXGGrammarDFAStateRelation class.

The method **resetDFAState()** reset the DFA state relation.

### 2.12. Class LXGGrammarLine

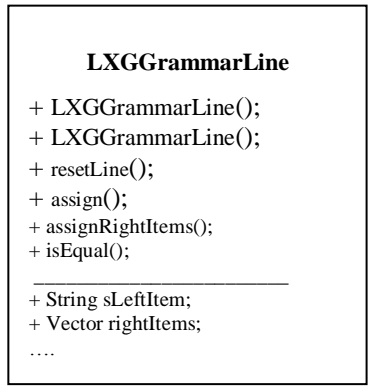

The class LXGGrammarLine (see LXGGrammarLine.java) represents the LXG grammar line structure used by the LXG Compiler. All the class members are public. The public class members make the use of the class easier. The class is for internal use only, and the public class members do not violate the class privacy and capsulation.

There are two overloaded constructors implemented within the LXGGrammarLine class. They differ by their parameters.

The method **resetLine()** resets all the class members to their initial state.

The method **assign()** assigns a LXGGrammarLine to the current one.

The method **assignRightItems()** assigns the right grammar items from a grammar line to the current one.

The method **isEqual()** checks if the grammar line is equal to another one.

### 2.13. Class LXGGrammarSet

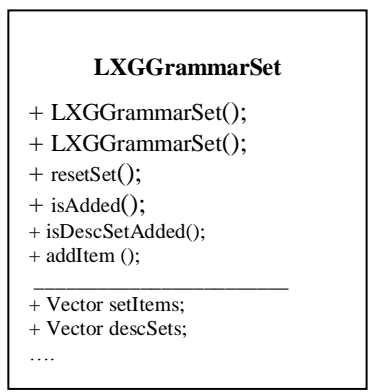

The class LXGGrammarSet (see LXGGrammarSet.java) represents the LXG grammar set structure used by the LXG Compiler. All the class members are public. The public class members make the use of the class easier. The class is for internal use only, and the public class members do not violate the class privacy and capsulation.

There are two overloaded constructors implemented within the LXGGrammarSet class. They differ by their parameters.

The method *resetSet()* resets all the class members to their initial state.

The method *isAdded()* checks if an item is been already added to the grammar set.

The method *addItem()* adds an item to the grammar set.

### 2.14. Class LXGExceptions

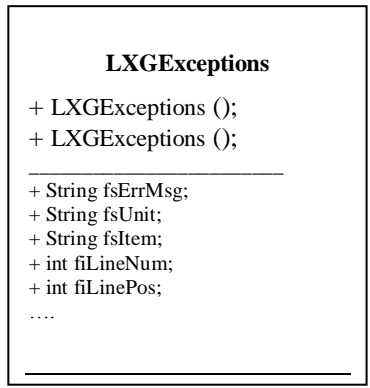

The class LXGExceptions (see LXGExceptions.java) represents the LXG exception structure used by the LXG Compiler. All the class members are public. The public class members make the use of the class easier. The class is for internal use only, and the public class members do not violate the class privacy and capsulation. The class of LXGExceptions (see LXGExceptions.java) implements the super class for all the exceptions used and going to be used by the LXG Compiler.

There are two overloaded constructors implemented within the LXGExceptions class. They differ by their parameters.

The class member *fsErrMsg* keeps the error message.

The class member *fsUnit* keeps the compiler unit, which has raised the error.

The class member *fsItem* keeps the item name (usually it is a token name), which has caused the error.

The class member *fiLineNum* keeps the line number of the source code where the *fsItem* has been found.

The class member *fiLinePos* keeps the position of the source code where the *fsItem* has been found.

### 2.15. Class LXGGlobal

The class of LXGGlobal (see LXGGlobal.java) implements all the definitions of token names, special symbols, file names and other useful semantics used by whole project. This class acts a s a container for all the global variables(objects) used by the LXG Compiler.

### 2.16. Class ExceptionIncorrectScannerState

The class of ExceptionIncorrectScannerState (see LXGScanner.java) implements the exception for incorrect scanner state. This class derives from LXGExceptions and is used only by LXGScanner.

## 3. Testing and Outputs

Several tests have been performed to prove the correctness of the different LXG Compiler units. The units have been tested separately and together.

### 3.1. Scanner tests

### 3.1.1. Single-character symbol-tokens recognition

| Source input | Scan output |
|---|---|
| | bof |
| + | 1: PLUS, lexeme -> + |
| - | 2: MINUS, lexeme -> - |
| * | 3: MULTI, lexeme -> * |
| / | 4: OVER, lexeme -> / |
| ^ | 5: POWER, lexeme -> ^ |
| , | 6: COMMA, lexeme -> , |
| ; | 7: SEMI, lexeme -> ; |
| < | 8: LESS, lexeme -> < |
| > | 9: BIGGER, lexeme -> > |
| = | 10: EQUAL, lexeme -> = |
| # | 11: DIFF, lexeme -> # |
| ( | 12: LPAREN, lexeme -> ( |
| ) | 13: RPAREN, lexeme -> ) |
| [ | 14: LSQPAREN, lexeme -> [ |
| ] | 15: RSQPAREN, lexeme -> ] |
| | eof |

### 3.1.2. Multiple-character symbol-tokens recognition

| Source input | Scan output |
|---|---|
| := <br> :=: <br> <= <br> >= <br> ,..., | bof <br> 1: ASSIGN, lexeme -> := <br> 2: SWAP, lexeme -> :=: <br> 3: LESSEQ, lexeme -> <= <br> 4: BIGEQ, lexeme -> >= <br> 5: ELLIPSIS, lexeme -> ,..., <br> eof |

### 3.1.3. Reserved word tokens recognition

| Source input | Scan output |
|---|---|
| AND <br> DO <br> IF <br> OR <br> ARRAY <br> ELSE <br> INTEGER <br> PROCEDURE <br> VALUE <br> BEGIN <br> END <br> MOD <br> REM <br> WHILE <br> BOOLEAN <br> FOR <br> NOT <br> STRING | bof <br> 1: reserved word -> AND <br> 2: reserved word -> DO <br> 3: reserved word -> IF <br> 4: reserved word -> OR <br> 5: reserved word -> ARRAY <br> 6: reserved word -> ELSE <br> 7: reserved word -> INTEGER <br> 8: reserved word -> PROCEDURE <br> 9: reserved word -> VALUE <br> 10: reserved word -> BEGIN <br> 11: reserved word -> END <br> 12: reserved word -> MOD <br> 13: reserved word -> REM <br> 14: reserved word -> WHILE <br> 15: reserved word -> BOOLEAN <br> 16: reserved word -> FOR <br> 17: reserved word -> NOT <br> 18: reserved word -> STRING <br> eof |

### 3.1.4. Other tokens recognition

| Source input | Scan output |
|---|---|
| INTEGER X23; | bof<br>1: reserved word -> INTEGER<br>1: ID, lexeme -> X23<br>1: SEMI, lexeme -> ; |
| STRING ST; | 2: reserved word -> STRING<br>2: ID, lexeme -> ST<br>2: SEMI, lexeme -> ; |
| BOOLEAN B; | 3: reserved word -> BOOLEAN<br>3: ID, lexeme -> B<br>3: SEMI, lexeme -> ; |
| X23 := 4567; | 4: ID, lexeme -> X23<br>4: ASSIGN, lexeme -> :=<br>4: NUMBER, lexeme -> 4567, size = 4<br>4: SEMI, lexeme -> ; |
| ST := 'dddFFF"fff'; | 5: ID, lexeme -> ST<br>5: ASSIGN, lexeme -> :=<br>5: STRING, lexeme -> dddFFF"fff, size = 10<br>5: SEMI, lexeme -> ; |
| B:= TRUE; | 6: ID, lexeme -> B<br>6: ASSIGN, lexeme -> :=<br>6: BOOLEAN, lexeme -> TRUE<br>6: SEMI, lexeme -> ;<br>eof |

### 3.1.5. Error trapping

| Source input | Scan output |
|---|---|
| INTEGER K, I, N; | bof<br>1: reserved word -> INTEGER<br>1: ID, lexeme -> K<br>1: COMMA, lexeme -> ,<br>1: ID, lexeme -> I<br>1: COMMA, lexeme -> ,<br>1: ID, lexeme -> N<br>1: SEMI, lexeme -> ; |
| FOR K:=1,3,...,19 DO | 2: reserved word -> FOR<br>2: ID, lexeme -> K<br>2: ASSIGN, lexeme -> :=<br>2: NUMBER, lexeme -> 1, size = 1<br>2: COMMA, lexeme -> , |
| FOR I:=1,3,..,19 DO | 2: NUMBER, lexeme -> 3, size = 1 |
| | 2: ELLIPSIS, lexeme -> ,..., |
| | 2: NUMBER, lexeme -> 19, size = 2 |
| | 2: reserved word -> DO |
| | 3: reserved word -> FOR |
| | 3: ID, lexeme -> I |
| | 3: ASSIGN, lexeme -> := |
| | 3: NUMBER, lexeme -> 1, size = 1 |
| | 3: COMMA, lexeme -> , |
| | 3: NUMBER, lexeme -> 3, size = 1 |
| | 3: error, incorrect token -> ,.. |
| | 3: COMMA, lexeme -> , |
| | 3: NUMBER, lexeme -> 19, size = 2 |
| | 3: reserved word -> DO |
| N := N + K + I; | 4: ID, lexeme -> N |
| | 4: ASSIGN, lexeme -> := |
| | 4: ID, lexeme -> N |
| | 4: PLUS, lexeme -> + |
| | 4: ID, lexeme -> K |
| | 4: PLUS, lexeme -> + |
| | 4: ID, lexeme -> I |
| | 4: SEMI, lexeme -> ; |
| N:= 12345678901; | 5: ID, lexeme -> N |
| | 5: ASSIGN, lexeme -> := |
| | 5: error, number size over 10 digits -> 12345678901, size = 11 |
| | 5: SEMI, lexeme -> ; |
| STRING S, S:, St; | 6: reserved word -> STRING |
| | 6: ID, lexeme -> S |
| | 6: COMMA, lexeme -> , |
| | 6: ID, lexeme -> S |
| | 6: error, incorrect token -> : |
| | 6: COMMA, lexeme -> , |
| | 6: ID, lexeme -> S |
| | 6: error, incorrect token -> t |
| | 6: SEMI, lexeme -> ; |
| | eof |

### 3.2. Grammar scanner tests

#### 3.2.1. Computing First sets

*First set:* the set of terminals, which can start any derivation from N.

First(**program**) = { bof }
First(**prgm-body**) = { PROCEDURE , BEGIN , FOR , WHILE , IF , REM , bIdentifier ,
      sIdentifier , aIdentifier , iIdentifier , uIdentifier }
First(**stmnt-list**) = { BEGIN , FOR , WHILE , IF , REM , bIdentifier , sIdentifier ,
      aIdentifier , iIdentifier , uIdentifier }
First(**stmnt**) = { BEGIN , FOR , WHILE , IF , REM , bIdentifier , sIdentifier , aIdentifier
      , iIdentifier , uIdentifier }
First(**for-list**) = { + , - , ( , number , aIdentifier , iIdentifier }
First(**for-item**) = { + , - , ( , number , aIdentifier , iIdentifier }
First(**proc-call**) = { uIdentifier }
First(**exp-list**) = { string , aIdentifier , sIdentifier , + , - , NOT , ( , number , boolean ,
      bIdentifier , iIdentifier }
First(**exp-item**) = { string , aIdentifier , sIdentifier , + , - , NOT , ( , number , boolean ,
      bIdentifier , iIdentifier }
First(**int-exp**) = { + , - , ( , number , aIdentifier , iIdentifier }
First(**int-term**) = { + , - , ( , number , aIdentifier , iIdentifier }
First(**int-fact**) = { + , - , ( , number , aIdentifier , iIdentifier }
First(**int-prim**) = { ( , number , aIdentifier , iIdentifier }
First(**int-dest**) = { aIdentifier , iIdentifier }
First(**int-ident**) = { iIdentifier }
First(**bool-exp**) = { NOT , ( , boolean , bIdentifier , + , - , number , aIdentifier , iIdentifier
      }
First(**bool-term**) = { NOT , ( , boolean , bIdentifier , + , - , number , aIdentifier ,
      iIdentifier }
First(**bool-fact**) = { NOT , ( , boolean , bIdentifier , + , - , number , aIdentifier , iIdentifier
      }
First(**bool-prim**) = { ( , boolean , bIdentifier , + , - , number , aIdentifier , iIdentifier }
First(**bool-reln**) = { + , - , ( , number , aIdentifier , iIdentifier }
First(**bool-ident**) = { bIdentifier }
First(**str-exp**) = { string , sIdentifier }
First(**str-ident**) = { sIdentifier }
First(**arr-ident**) = { aIdentifier }
First(**ident-list**) = { iIdentifier , bIdentifier , sIdentifier , aIdentifier }
First(**ident-item**) = { iIdentifier , bIdentifier , sIdentifier , aIdentifier }
First(**proc-decl**) = { PROCEDURE }
First(**proc-head**) = { uIdentifier }
First(**proc-ident**) = { uIdentifier }

### 3.2.2. Computing Follow sets

*Follow set:* the set of terminals, which can follow a specific nonterminal in a rightmost derivation.

Follow(**program**) = { $ }
Follow(**prgm-body**) = { eof }
Follow(**stmnt-list**) = { ; , END , eof }
Follow(**stmnt**) = { ELSE , ; , END , eof }
Follow(**for-list**) = { DO , , }
Follow(**for-item**) = { DO , , }
Follow(**proc-call**) = { ELSE , ; , END , eof }
Follow(**exp-list**) = { ) , , }
Follow(**exp-item**) = { ) , , }
Follow(**int-exp**) = { / , ,...., + , - , ) , ] , < , <= , = , >= , > , # , MOD , ELSE , ; , END , eof , DO , , , AND , THEN , OR }
Follow(**int-term**) = { * , / , ,...., + , - , ) , ] , < , <= , = , >= , > , # , MOD , ELSE , ; , END , eof , DO , , , AND , THEN , OR }
Follow(**int-fact**) = { * , / , ,...., + , - , ) , ] , < , <= , = , >= , > , # , MOD , ELSE , ; , END , eof , DO , , , AND , THEN , OR }
Follow(**int-prim**) = { ^ , * , / , ,...., + , - , ) , ] , < , <= , = , >= , > , # , MOD , ELSE , ; , END , eof , DO , , , AND , THEN , OR }
Follow(**int-dest**) = { := , REM , :=: , ELSE , ; , END , eof , ^ , * , / , ,...., + , - , ) , ] , < , <= , = , >= , > , # , MOD , DO , , , AND , THEN , OR }
Follow(**int-ident**) = { := , REM , :=: , ELSE , ; , END , eof , ^ , * , / , ,...., + , - , ) , ] , < , <= , = , >= , > , # , MOD , DO , , , AND , THEN , OR }
Follow(**bool-exp**) = { DO , THEN , OR , ) , ELSE , ; , END , eof , , }
Follow(**bool-term**) = { AND , DO , THEN , OR , ) , ELSE , ; , END , eof , , }
Follow(**bool-fact**) = { AND , DO , THEN , OR , ) , ELSE , ; , END , eof , , }
Follow(**bool-prim**) = { AND , DO , THEN , OR , ) , ELSE , ; , END , eof , , }
Follow(**bool-reln**) = { AND , DO , THEN , OR , ) , ELSE , ; , END , eof , , }
Follow(**bool-ident**) = { := , :=: , ELSE , ; , END , eof , AND , DO , THEN , OR , ) , , }
Follow(**str-exp**) = { ELSE , ; , END , eof , ) , , }
Follow(**str-ident**) = { := , :=: , ELSE , ; , END , eof , ) , , }
Follow(**arr-ident**) = { [ , ) , , }
Follow(**ident-list**) = { ) }
Follow(**ident-item**) = { , , ) }
Follow(**proc-decl**) = { ; }
Follow(**proc-head**) = { ; }
Follow(**proc-ident**) = { ( , ELSE , ; , END , eof }

### 3.2.3. Determining the number of all the DFA states

**158** DFA states

### 3.2.4. Determining the number of the DFA states with a shift-reduce conflict

Only one DFA state with a shift-reduce conflict has been found :

Reduce transitions: Follow(stmnt) : 8,
Shift transitions: ELSE : 144,
Goto transitions:

**stmnt -> IF bool-exp THEN stmnt .**
**stmnt -> IF bool-exp THEN stmnt .ELSE stmnt**

### 3.3. Pre-parser tests

The Pre-parser has been tested thoroughly with all the provided examples and additional tests. The additional tests have checked the correctness of variable overlapping and parameters declaration. For additional tests run the LXG Compiler in a "dump to file" mode and check the file "lxg_symbol_table.out" for the generated symbol table. The following is a simple test, showing the correctness of the Pre-parser.

LXG code:
*INTEGER A,B;*
*ARRAY AA[20];*

*FOR A:= 1,...,A,B,...,20 DO WRITES("A");*

*AA[2]:= B*

Symbol table generated by the LXG Pre-parser:

name: A; type: INTEGER; is_generated: false; location: MAIN_PROGRAM; scope: GLOBAL; is_param: false; is_value: false; lines: 6, 6,
name: B; type: INTEGER; is_generated: false; location: MAIN_PROGRAM; scope: GLOBAL; is_param: false; is_value: false; lines: 6, 8,
name: AA; type: ARRAY; size: 20; is_generated: false; location: MAIN_PROGRAM; scope: GLOBAL; is_param: false; is_value: false; lines: 8,
name: I_00001; type: INTEGER; is_generated: true; value: ; location: MAIN_PROGRAM; scope: GLOBAL; lines: 6
name: I_00002; type: INTEGER; is_generated: true; value: ; location: MAIN_PROGRAM; scope: GLOBAL; lines: 6
name: S_00001; type: STRING; is_generated: true; value: ; location: MAIN_PROGRAM; scope: GLOBAL; lines: 6
name: I_00003; type: INTEGER; is_generated: true; value: ; location: MAIN_PROGRAM; scope: GLOBAL; lines: 8

### 3.4. Parser tests

The Parser has been tested thoroughly with all the provided examples and additional tests. The additional tests have been performed mainly for error trapping check. To perform more additional tests run the LXG Compiler in a "dump to file" mode and check the file "lxg_parse.out" for the generated parser out. The following is a simple test, showing the correctness of the Parser.

LXG code:
>    *INTEGER A,B;*
>    *A:= A+B*

Out generated by the LXG Parser:

A : int-ident -> iIdentifier
A : int-dest -> int-ident
A : int-ident -> iIdentifier
A : int-dest -> int-ident
A : int-prim -> int-dest
A : int-fact -> int-prim
A : int-term -> int-fact
A : int-exp -> int-term
B : int-ident -> iIdentifier
B : int-dest -> int-ident
B : int-prim -> int-dest
B : int-fact -> int-prim
B : int-term -> int-fact
B : int-exp -> int-exp + int-term
B : stmnt -> int-dest := int-exp
B : stmnt-list -> stmnt
B : prgm-body -> stmnt-list
B : program -> bof prgm-body eof

### 3.5. Code Generator tests

The Code Generator has been tested thoroughly with all the provided examples and additional tests. The additional tests have been performed mainly for an error trapping check (trapping the run-time errors). Because the Code Generator is the last unit from the LXG Compiler "chain" by testing it we test the entire compiler.

LXG code:
>        **INTEGER A,B;**
>        **A:= A+B**

## Assembler code generated by LXGCompiler:

```
%%%% =========================================================%
%%%% test00.m
%%%%
%%%% Compiled with LXGCompiler, developed by Emil Vassev
%%%% =========================================================%

%----------------- program's entry point -------------------%
           entry

           jl    R14,Sy_Init_SP

           lw    R1,A(R0)
           lw    R2,B(R0)
           add   R2,R1,R2
           sw    A(R0),R2

           j     LB_EXIT

%===================== RUN-TIME ERROR TRAPS ===================%

%---- run-time error trap: Recursive function call -----------%
 LB_RECURSIVE_CALL
           jl    R14,Sy_Push
           dw    ERR_RECURS_CALL
           jl    R14,WRITES
           j     LB_EXIT

%---- run-time error trap: FOR-DO zero denominator -----------%
 LB_FOR_DO_ZERO_DEN
           jl    R14,Sy_Push
           dw    ERR_ZERO_DENOM
           jl    R14,WRITES
           j     LB_EXIT

%---- run-time error trap: array zero-index writing ----------%
 LB_ARRAY_ZERO_INDEX
           jl    R14,Sy_Push
           dw    ERR_ZERO_INDEX
           jl    R14,WRITES
           j     LB_EXIT

%---- run-time error trap: array low boundary ----------------%
 LB_ARRAY_LOW_BOUND
           jl    R14,Sy_Push
           dw    ERR_BOT_BOUNDERY
           jl    R14,WRITES
           j     LB_EXIT

%---- run-time error trap: array up boundary -----------------%
 LB_ARRAY_UP_BOUND
           jl    R14,Sy_Push
           dw    ERR_TOP_BOUNDERY
           jl    R14,WRITES
```

```
                    j    LB_EXIT

%---- run-time error trap: division by zero ------------------%
  LB_DIV_ZERO
                    jl   R14,Sy_Push
                    dw   ERR_DIV_ZERO
                    jl   R14,WRITES
                    j    LB_EXIT

%---- run-time error trap: negative exponent -----------------%
  LB_NEG_EXPONENT
                    jl   R14,Sy_Push
                    dw   ERR_NEG_EXPONENT
                    jl   R14,WRITES
                    j    LB_EXIT

  LB_EXIT
                    hlt

%====================== var declarations ====================%

                    align
  READC_Y           res  4
  READN_Y           res  4
  WRITEC_X          res  4
  WRITEN_X          res  4
  WRITEN_H          res  4
  WRITES_S          res  4
  SPACE_X           res  4
  LINE_X            res  4
  A         res  4
  B         res  4

%================= registry save variables ==================%
%---- procedures ----%
%---- FOR ... DO ----%

%================ run-time errors definition ================%
  INT_VAL_ZERO      dw   0
  INT_VAL_ONE       dw   1
  INT_VAL_ANY       dw   1
  ERR_DIV_ZERO      dw   ERR_DIV_ZERO_A
  ERR_DIV_ZERO_A    db   "LXG run-time error: Division by zero",0
                    align
  ERR_NEG_EXPONENT  dw   ERR_NEG_EXPONENT_A
  ERR_NEG_EXPONENT_A db   "LXG run-time error: Negative exponent",0
                    align
  ERR_BOT_BOUNDERY  dw   ERR_BOT_BOUNDERY_A
  ERR_BOT_BOUNDERY_A db   "LXG run-time error: Array index out of range (index < 0)",0
                    align
  ERR_TOP_BOUNDERY  dw   ERR_TOP_BOUNDERY_A
  ERR_TOP_BOUNDERY_A db   "LXG run-time error: Array index out of range (index > top)",0
                    align
  ERR_ZERO_INDEX    dw   ERR_ZERO_INDEX_A
  ERR_ZERO_INDEX_A  db   "LXG run-time error: Writing to the zero-index element of an array",0
                    align
  ERR_ZERO_DENOM    dw   ERR_ZERO_DENOM_A
  ERR_ZERO_DENOM_A  db   "LXG run-time error: A zero denominator used by FOR-DO loop",0
                    align
  ERR_RECURS_CALL   dw   ERR_RECURS_CALL_A
  ERR_RECURS_CALL_A db   "LXG run-time error: Recursive function call",0
                    align
%===========================================================%
```

## Appendix A: Grammar for LXG

program → **bof** prgm-body **eof**
prgm-body → proc-decl **;** prgm-body | stmnt-list
stmnt-list → stmnt | stmnt-list **;** stmnt
stmnt → **BEGIN** stmnt-list **END** |
    **FOR** int-dest := for-list **DO** stmnt |
    **WHILE** bool-exp **DO** stmnt |
    **IF** bool-exp **THEN** stmnt |
    **IF** bool-exp **THEN** stmnt **ELSE** stmnt |
    int-dest := int-exp |
    int-dest := int-exp **/** int-exp |
    **REM** int-dest := int-exp **/** int-exp |
    int-dest **REM** int-dest := int-exp **/** int-exp |
    bool-ident := bool-exp |
    str-ident := str-exp |
    int-dest :=: int-dest |
    bool-ident :=: bool-ident |
    str-ident :=: str-ident |
    proc-call
for-list → for-item | for-list **,** for-item
for-item → int-exp | int-exp **,...,** int-exp
proc-call → proc-ident | proc-ident **(** exp-list **)**
exp-list → exp-item | exp-list **,** exp-item
exp-item → int-exp | bool-exp | str-exp | arr-ident
int-exp → int-term | int-exp **+** int-term | int-exp **-** int-term
int-term → int-fact | int-term **\*** int-fact
int-fact → int-prim | int-prim **^** int-fact | **+** int-fact | **-** int-fact
int-prim → int-dest | **(** int-exp **)** | *number*
int-dest → arr-ident **[** int-exp **]** | int-ident
int-ident → *iIdentifier*
bool-exp → bool-term | bool-exp **OR** bool-term
bool-term → bool-fact | bool-term **AND** bool-fact
bool-fact → bool-prim | **NOT** bool-fact
bool-prim → bool-ident | bool-reln | **(** bool-exp **)** | *boolean*
bool-reln → int-exp **<** int-exp | int-exp **<=** int-exp |
    int-exp **=** int-exp | int-exp **>=** int-exp |
    int-exp **>** int-exp | int-exp **#** int-exp |
    int-exp **=** int-exp **MOD** int-exp |
    int-exp **#** int-exp **MOD** int-exp
bool-ident → *bIdentifier*
str-exp → str-ident | *string*
str-ident → *sIdentifier*
arr-ident → *aIdentifier*
ident-list → ident-item | ident-item **,** ident-list
ident-item → int-ident | bool-ident | str-ident | arr-ident
proc-decl → **PROCEDURE** proc-head **;** stmnt-list **END**
proc-head → proc-ident | proc-ident **(** ident-list **)**
proc-ident → *uIdentifier*